\newcommand{\bfactor}{\beta\textrm{-factor}}
\newcommand{\gc}{\gamma_{p}}
\newcommand{\gr}{\gamma_r}
\newcommand{\Rth}{R_t}
\newcommand{\Pth}{P_t}

\documentclass[12pt]{iopart}
\usepackage{iopams}

\usepackage{graphicx,epsfig}

\begin{document}

\title[High-cooperativity nanofiber laser]{High-cooperativity nanofiber laser}

\author{Sanli Faez$^{*}$, Pierre T\"urschmann, and Vahid Sandoghdar}
\address{ Max Planck Institute for the Science of Light (MPL), 91058 Erlangen, Germany}
\address{ Friedrich Alexander University, 91058 Erlangen, Germany}
\address{$^{*}$ Current address: Huygens-Kamerlingh Onnes Laboratory, Leiden Institute of Physics, Leiden University, PO Box 9504, 2300 RA Leiden, The Netherlands}
\eads{\mailto{faez@physics.leidenuniv.nl}, \mailto{vahid.sandoghdar@mpl.mpg.de}}


\begin{abstract}
Cavity-free efficient coupling between emitters and guided modes is of great current interest for nonlinear quantum optics as well as efficient and scalable quantum information processing. In this work, we extend these activities to the coupling of organic dye molecules to a highly confined mode of a nanofiber, allowing mirrorless and low-threshold laser action in an effective mode volume of less than 100 femtoliters. We model this laser system based on semi-classical rate equations and present an analytic compact form of the laser output intensity. Despite the lack of a cavity structure, we achieve a coupling efficiency of the spontaneous emission to the waveguide mode of $\beta = 0.07 \pm 0.01$, in agreement with our calculations. In a further experiment, we also demonstrate the use of a plasmonic nanoparticle as a dispersive output coupler. Our laser architecture is promising for a number of applications in optofluidics and provides a fundamental model system for studying nonresonant feedback stimulated emission.  \end{abstract}


\maketitle

Miniature light sources that can be integrated in chip-scale devices are one of the fundamental building blocks of photonics circuits. Among these, micro and nanolasers are particularly attractive because they offer a high degree of coherence, tunability, directionality, and efficiency. However, reducing the size of the gain medium often compromises the laser threshold and output stability. A customary strategy to counter these undesirable effects is to improve the resonator quality factor~\cite{sandoghdar_very_1996, painter_two-dimensional_1999, vahala_optical_2003}. An alternative solution is to maximize the so-called $\bfactor$ or cooperativity, which is defined as the fraction of the spontaneous emission into the optical mode that hosts the laser radiation relative to the total emission~\cite{yamamoto_microcavity_1991,van_exter_two_1996,altug_ultrafast_2006}. As $\beta$ approaches unity, the laser threshold is expected to vanish~\cite{yamamoto_microcavity_1991}. Many efforts have been pursued to increase $\beta$ via concepts from cavity quantum electrodynamics based on the modification of the density of states in resonant structures~\cite{yamamoto_microcavity_1991,khajavikhan_thresholdless_2012}, but the realization of suitable microcavity geometries often confronts limitations imposed by the choice of material and fabrication technologies~\cite{vahala_optical_2003}.

Recently, we have shown that unity coupling efficiency between light and matter can be achieved for propagating photons in a cavity-free single-pass geometry if the spatial mode of the photons and the emission pattern of the atoms are matched~\cite{agio_prl_2008}. Considering that the emission associated with a typical optical transition follows the radiation pattern of an electric dipole, this means that the photons have to be produced in a dipolar mode. Although the overlap between a tightly-focused light beam and a dipolar radiation mode is very high~\cite{mojarad_josa}, achieving unity efficiency in this strategy is nontrivial. As an alternative, however, one can engineer the near-field optical interactions to funnel the radiation of point-like emitters into more conventional directed light beams. Some of the avenues discussed in literature either use planar metallo-dielectric antennas~\cite{lee_planar_2011}, or linearly extended architectures such as co-axial~\cite{khajavikhan_thresholdless_2012}, photonic crystal~\cite{lund-hansen_experimental_2008}, or slot~\cite{quan_broadband_2010} waveguides, or plasmonic nanowires~\cite{chang_singlephoton_2007}, or simply nanofibers~\cite{shen_coherent_2005,rephaeli_stimulated_2012}. In the current work, we choose the latter geometry, which has been experimentally explored by coupling tapered fibers to atoms, molecules, or quantum dots~\cite{vetsch_optical_2010,goban_demonstration_2012,yalla_efficient_2012}. The $\bfactor$ in such waveguiding systems depends on the refractive index contrast and dimensions of the fiber as well as the coordinates of the emitter relative to the mode. Interestingly, lasing from a short semiconductor nanowire with substantial feedback from its endfacets has also been reported about a decade ago although the key issues of its operation were not explained~\cite{duan_single-nanowire_2003}.

In this article, we propose and examine the coupling of a gain medium to a nanofiber without an explicit use of any cavity geometry. As a flexible fabrication platform we use glass nanocapillaries, which can be easily filled with organic dye molecules, quantum dots, or other optically active materials. By choosing a filling medium with an index of refraction higher than glass, we can achieve a tightly-confined mode of the waveguide, which yields a maximum $\beta$ of 0.12 for emitters on the axis of the core. Furthermore, we show that individual gold nanoparticles can serve as efficient dispersive outcouplers. We discuss the theoretical and experimental performance of our nanofiber laser and its connection to nonresonant feedback laser (nRFL) action~\cite{ambartsumyan_laser_1967}, commonly referred to as random laser~\cite{wiersma_physics_2008}. Two of the central features of this lasing regime are intrinsic multimode operation and lack of well-defined cavity geometry. These have rendered a quantitative understanding of the laser operation both a theoretical and an experimental challenge. Reduction of the geometry to one dimension and single transverse lasing operation can substantially simplify the system. We compare our measurements with a simplified model based on semi-classical rate equations. Using an explicit and compact formula for the output laser power as a function of pump rate we can extract the $\bfactor$ from a direct fit to our measured data. Finally, our work has important implications for enhancing the performance of optofluidic fiber lasers, where $\bfactor$s have been typically very low either due to the large mode area in the waveguide or the presence of multiple transverse modes~\cite{yang_mirrorless_2000,vasdekis_fluidic_2007,li_optofluidic_2008}.

\section{Experimental methods}

Our laser architecture consists of a fused-silica glass capillary with an inner diameter of 710~nm filled with phenol and rhodomine-6G as gain medium. To fill the capillary, a short piece is immersed into a 4.7~mM dye solution kept in the liquid form at $T=50^{\circ}{\rm C} $. The solution rises inside the capillary for about 30~mm within 100 seconds. The capillary is then taken out and the excess material on the outer surfaces is wiped off with ethanol. Meanwhile, the phenol inside the capillary solidifies at room temperature. A 3-mm piece of the filled nanocapillary is cleaved and glued vertically on a glass cover slip with ultraviolet-curable glue (NOA61, Norland Optical Adhesives). The refractive index of this glue $(n=1.54)$ is matched with that of phenol.

\begin{figure}[t]
\includegraphics[width=16cm]{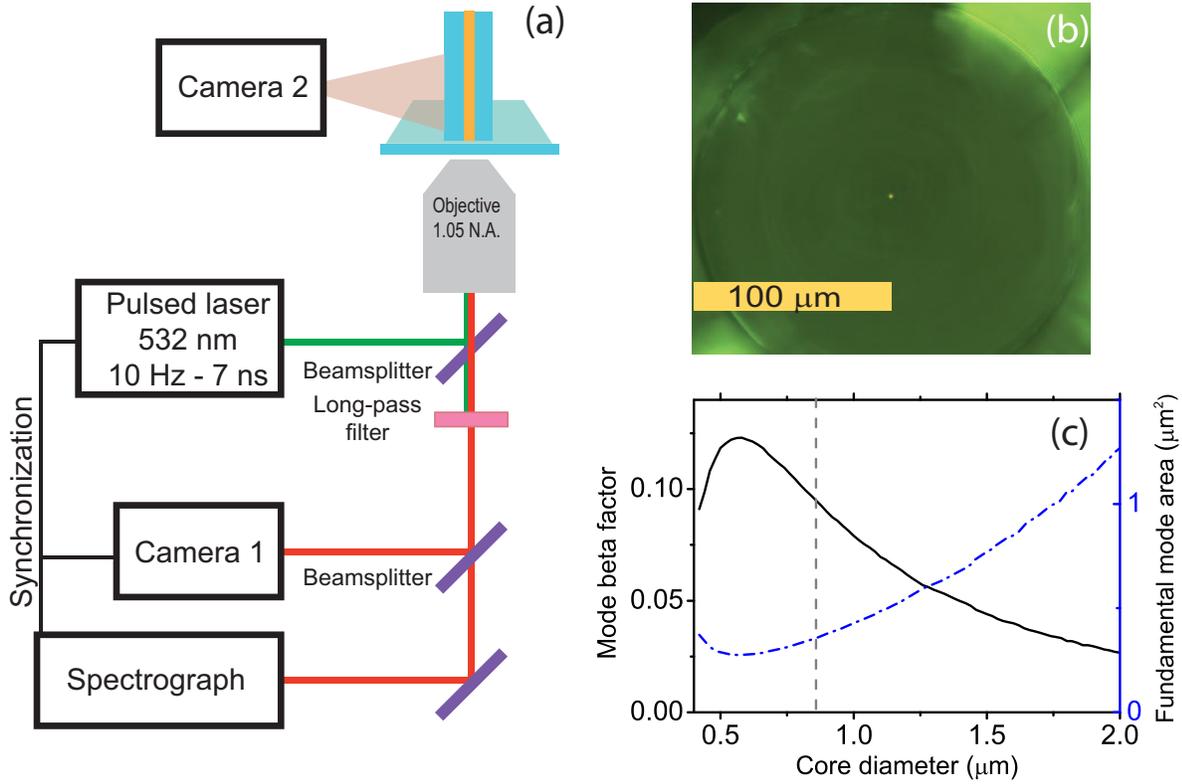}
\caption{\label{fig:fibertop}(a) Sketch of the measurement setup. The sample consists of a silica glass capillary with an inner diameter of 710~nm and an outer diameter of 200~$\mu$m. It is filled with a mixture of organic dye rhodomine-6G in phenol. (b) Passive microscope image from the cleaved end of the filled capillary fixed on a glass slide with index-matched glue. (c) Theoretical $\bfactor$ and fundamental-mode area for a cylindrical waveguide considering core and cladding refractive indices of $n_c=1.54$ and $n_o=1.45$. The vertical dashed line indicates the maximal diameter for single-mode operation at $\lambda_0=580$~nm. For larger diameters the total coupling of emission to the fiber increases because of the emergence of higher order modes, but the coupling to the fundamental mode decreases.}
\end{figure}

We determined the absorption length of the gain medium to be 37~$\mu$m by measuring, perpendicular to the fiber axis, the decay of the red-shifted fluorescence along the core, while exciting the guided mode with continuous-wave light at 532 nm at one end. The experimental setup is sketched in Fig.~\ref{fig:fibertop}(a). Since the gain medium extends further than its absorption length, any emission produced at the bottom end does not interact with the upper end of the active material. On the lower side, the medium is largely index-matched so that reflections from the capillary end should amount to less than $10^{-3}$. Furthermore, because the output of the nanocapillary mode with a numerical aperture of 0.46 is very divergent, reflections from secondary interfaces farther away are negligible so that we can consider the external feedback to the system to be very small. To pump the medium for laser action, light from a frequency-doubled Q-switched Nd-YAG laser with 10 Hz repetition rate and pulse duration of 7~ns was attenuated and focused on the end of the nanocapillary through a high-NA immersion objective. The red-shifted emission was collected via the same microscope objective and directed to a spectrograph and a digital camera in parallel.
Figure~\ref{fig:fibertop}(b) shows the lower end facet of the capillary.

\section{Theoretical considerations}
\subsection{Spontaneous emission $\bfactor$}
The $\bfactor$ is defined as the ratio between the spontaneous emission into the guided mode and the total emission~\cite{chu_spontaneous_1993}. In our system, because of the moderate index contrast between the core and cladding and an operation wavelength far from the mode cut-off, the effect of mode dispersion on the emission lifetime is negligible and, hence, the total emission is almost the same as in a homogeneous medium. Considering this approximation~\cite{hwang_dye_2011}, the $\bfactor$ can be written as
\begin{eqnarray}
\beta&=&\frac{3\lambda_0^2}{8\pi \epsilon(\mathbf{r}_0)}\frac{1}{A_e}.
\label{eq:betafactor}
\end{eqnarray}
Here $\lambda_0$ is the emission wavelength in vacuum and the effective mode area $A_e$ is defined as
\begin{eqnarray}
A_e=\frac{\int\epsilon(\mathbf{r})|E(\mathbf{r})|^2 d\mathbf{r}}{\epsilon(\mathbf{r}_0)|E(\mathbf{r}_0)|^2},
\label{eq:effmode}
\end{eqnarray}
where $\epsilon(\mathbf{r})$ and $E(\mathbf{r})$ are the dielectric constant and the electric field distribution of the mode, respectively and $\mathbf{r}_0$ denotes the position of the emitter. The integration is performed over the plane perpendicular to the propagation direction. For a cylindrical waveguide, the modes can be analytically calculated and are given by Bessel functions~\cite{chu_spontaneous_1993}. In our case, the higher refractive index of the phenol core ($n_c=1.54$) as compared to that of silica ($n_s=1.45$) results in a tightly-confined single guided mode. As seen in Fig.~\ref{fig:fibertop}(c), the core diameter was chosen to optimize the $\bfactor$.

\subsection{Semiclassical rate equations}
In this section, we consider a one-dimensional gain medium of finite length $L$ with perfectly-matched boundaries and introduce a phenomenological $\bfactor$. Unlike the common textbook description of lasers, a nRFL requires no well-separated passive cavity mode for operation. The dominant frequency selection mechanism is the emission spectrum of the gain medium accompanied by the fast decay of the amplified stimulated emission. Therefore, the typical decay rate of photons out of the gain region is determined by the optical path length and is given by $\gc \approx c/L$. This quantity amounts to $10^{11}-10^{13}$~Hz for typical microlasers, with or without a cavity structure, and is usually much larger than the upper state decay rate of the gain medium ($\gc \gg \gr$).

Following a standard semiclassical picture~\cite{siegman_lasers_1986,rice_photon_1994}, it is possible to describe this system with simplified rate equations for the number of photons $q(t)$ present in the gain region and the number of molecules in the upper laser level $N(t)$
\begin{eqnarray}\label{eq:rateeqs}
\frac{dq}{dt}&=& \beta \gr (q+1) N -\gc q,\\ \nonumber
\frac{dN}{dt}&=& R - \beta \gr q N - \gr N .
\end{eqnarray}
Here, the effective pump rate $R$ is defined as
\begin{eqnarray}\label{eq:satpow}
R&\equiv& \frac{\eta}{h\nu}\frac{P}{1+P/P_s},
\end{eqnarray}
and quantities $P$, $\eta$ and $h\nu$ are the actual power, the overall coupling efficiency of the pump beam and  the photon energy, respectively. The pump saturation intensity $P_s$ is directly proportional to the concentration of dye molecules because population inversion can never exceed the total number of dye molecules in the system. This results in the transparency of the gain medium to the pump beam at high excitation rates. Equations~(\ref{eq:rateeqs}) and (\ref{eq:satpow}) describe an optically pumped 4-level gain medium with a very fast relaxation of the lower laser level to the ground state such that the population inversion equals the upper state occupancy at all times. In this treatment, spontaneous emission is included in a phenomenological fashion~\cite{rice_photon_1994}. This formalism has previously been used for extracting the quantum efficiency of the gain medium from the random laser emission output~\cite{el-dardiry_classification_2011}.

The analytic form of the photon occupancy of the laser mode as a function of pump rate is given by
\begin{eqnarray}
q&=&\frac{R}{2\gc}-\frac{1}{2\beta}+\frac{1}{2}\sqrt{\left(\frac{R}{\gc}-\frac{1}{\beta}\right)^2+4\frac{R}{\gc}}.
\label{eq:photons}
\end{eqnarray}
At elevated pump rates, but in the absence of absorption transparency, a linear relation between $q$ and $R$ is recovered. The intersection of this line with the $q=0$ axis defines the laser threshold pump rate and reads
\begin{eqnarray}
\Rth=(\frac{1}{\beta}-1)\gc.
\label{eq:threshold}
\end{eqnarray}
This minimal model correctly shows that the laser threshold is reduced by enhancing the $\bfactor$ and/or lowering $\gc$. In practice, changing these two parameters relates to distinct physical aspects of the system. The reduction of the cavity decay rate usually requires implementation of better resonators, which has also been the common approach for lowering the threshold of fluidic lasers~\cite{song_low-order_2009}. In this article, we take an alternative route and demonstrate how to realize a dye laser with higher $\bfactor$.

\section{Results}
\begin{figure}[t]	
\includegraphics[width=16cm]{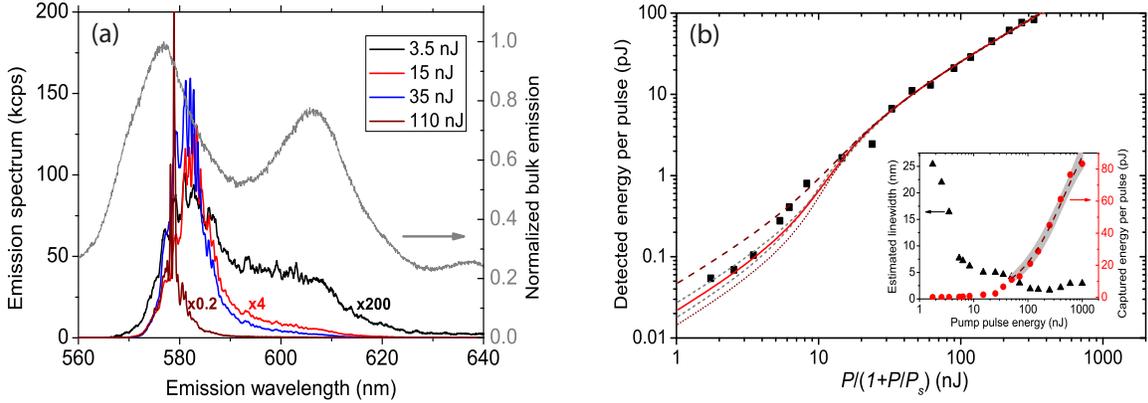}
\caption{\label{fig:threshold} (a) Emission spectra at higher pump pulse energies show line narrowing, relative to the bulk spectrum (grey line). The spectra at 3.5, 15 and 110 nJ are multiplied by factors 200, 4, and 0.2, respectively, for visibility. The bulk spectrum is measured by continuous-wave excitation of the same mixture inside a capillary with 20~$\mu$m inner diameter. (b) The emitted power, collected from the end of capillary, is plotted versus the rescaled pump power, which corrects for the  saturation of pump absorption. By fitting the scaled data points to Eq.~(\ref{eq:photons}), we find $\beta=0.07 \pm 0.01$ and $P=10.5 \pm1.2$~nJ/pulse for the laser threshold. The grey curves depict the fits for the higher and lower error margins of $\beta$. For comparison, the same fitting procedure is applied for imposed values of $\beta=0.05$ and $\beta=0.15$, respectively, and depicted in brown dotted and dashed lines. In the inset the measured total emission (red dots) and its spectral width (black triangles) are plotted as a function of measured pump power. The saturation power is determined to be $P_s = 485 \pm 30$~nJ by fitting to the data points far above laser threshold, depicted by the solid line. The grey shaded area depicts its 95\% confidence interval.}
\end{figure}

Single-shot emission spectra were recorded synchronously with the pump pulses at different excitation intensities. Figure~\ref{fig:threshold}(a) presents examples of the recorded spectra overlaid on the bulk emission spectrum under excitation with a continuous-wave laser at 532 nm. The resolution of the spectrometer was 0.1 nm. At pulse energies as low as 5~nJ, the width of the spectrum sharply decreased from 25 to 8 nm. The narrowing saturates at a minimal width of roughly 2 nm, as measured by fitting the spectra to Gaussian curves. The laser output is linearly polarized with a ratio $I_p/(I_p+I_{np})$ better than 0.9, where $I_p$ and $I_{np}$ are the polarized and unpolarized components of the emission, respectively. Using Fourier imaging, we checked that the emission is confined to the fundamental transverse mode for laser emission above threshold. We also observe a fine structure in the emission spectra, which we discuss below.

Figure~\ref{fig:threshold}(b) displays the total integrated emission power and its spectral width as a function of the pump pulse energy. We applied a two-step fitting procedure to obtain the pump power at the laser threshold, saturation power, and $\bfactor$. First, a saturation intensity of $P_s = 485 \pm 30$~nJ was determined by fitting the experimental data points obtained at pump powers above 80 nJ with a s-curve following Eq.~(\ref{eq:satpow}). We estimated the error by repeating the fitting procedure after removing each data point and considering the statistical fluctuations. This fit is presented in the inset of Fig.~\ref{fig:threshold}(b). Using the outcome for $P_s$ we rescaled the pump energies to obtain the effective pump rate. Next, we fitted all data points to Eq.~(\ref{eq:photons}) with rescaled pump rates, taking $\Rth$, $\beta$ and the collection efficiency as fitting parameters. The result indicates a robust laser threshold of $\Pth=10.5 \pm 1.2$ nJ/pulse and $\beta=0.07 \pm 0.01$. These errors were estimated based on repeating the fitting procedure while varying $P_s$ within its range of uncertainty. The results are overlaid on the rescaled data points in Fig.~\ref{fig:threshold}(b). The resulting value of the $\bfactor$ is in good agreement with our numerical estimation within the error limits. To emphasize the sufficient statistical significance of our data, the best fits for manually fixed values of $\beta=0.05$ and $\beta=0.15$ are also presented in the same graph. The emission intensity scales linearly with the effective pump rate below and well above the laser threshold. The kink in the intermediate pump rates is attributed to the amplified spontaneous emission~\cite{cao_pra_numeric,khajavikhan_thresholdless_2012}.

To verify the importance of a highly-confined guided mode, we filled the nanocapillary with ethylene glycol instead of phenol while keeping the same dye concentration. In this case, the refractive index of 1.44 for ethylene glycol is slightly less than that of the silica glass cladding, so that there is no guided mode in the core. Indeed, we obtained no laser action in the control sample even at pulse energies 400 times larger than the laser threshold value with phenol. We also note in passing that the obtained saturation power, which was the optimal compromise between brightness and efficiency, is well below the damage threshold of the devices filled with phenol at roughly 10~$\mu$J per pulse. Finally, we point out that previous experiments have also investigated mirrorless lasing in capillaries. For example, Vasdekis et al. demonstrated a threshold of 100 nJ/pulse for capillaries filled with perylene red dye doped in toluene~\cite{vasdekis_fluidic_2007}. However, having worked with capillaries of diameters as large as 2-$\mu$m and a lower index contrast of $n_s-n_c=0.04$, the $\bfactor$ in that report was ten times smaller than ours.

\subsection{Coupling to gold nanoparticles}

So far, we have demonstrated laser action facilitated by the high coupling efficiency to a confined guided mode without any explicit feedback mirror, outcoupler or passive resonances. For many photonic applications, however, controlled outcouplers and frequency-selective elements are desirable~\cite{li_optofluidic_2008}. The high $\bfactor$ of our nanoscopic system lends itself to the use of nanoparticles as mirrors and outcouplers because such Rayleigh scatterers also radiate like point dipoles. In particular, nanoparticles with dispersive spectral response, as is the case for plasmonic particles, would conveniently act as wavelength-selective elements. Here, we report on our first results in integrating gold nanoparticles in a nanofiber laser.

\begin{figure}[t]
\includegraphics[width=16cm]{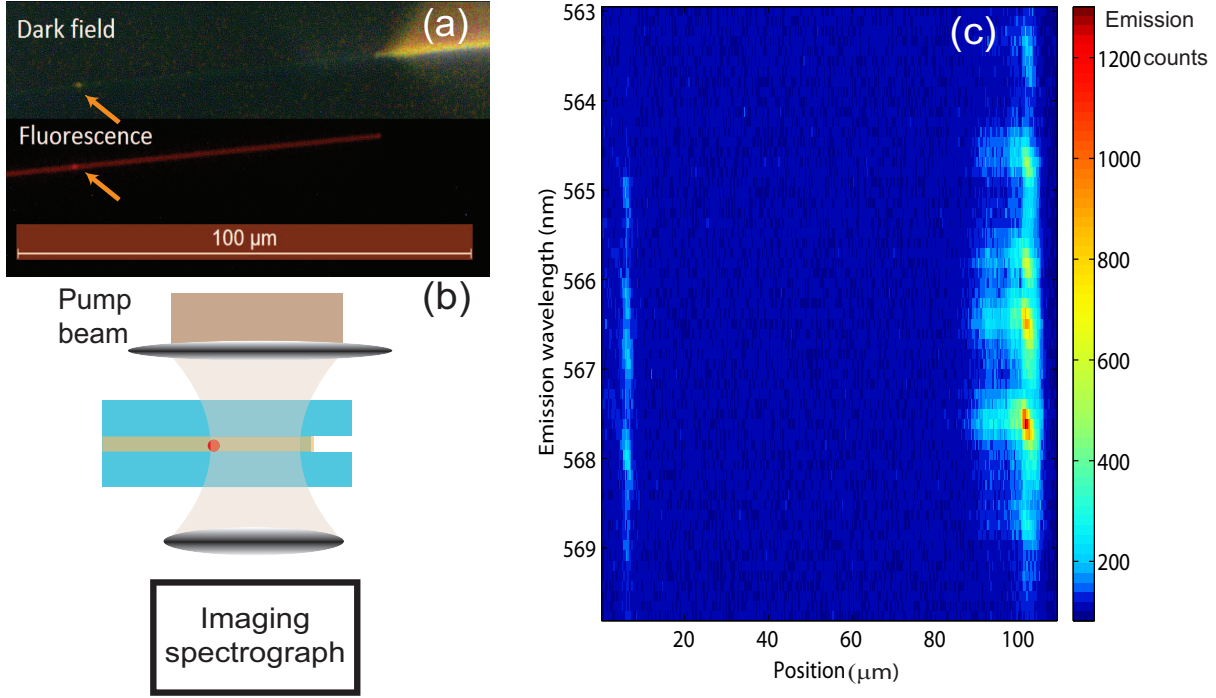}
\caption{\label{fig:fiberside}(a) Exemplary passive microscope images of a dye-filled fiber with one gold nanoparticle inside, captured in both dark-field and fluorescence modes. The gold nanoparticle is marked with arrows. (b) For recording the laser spectrum in this configuration, the capillaries were pumped from the side through a low-NA lens. The emission was collected through a second high-NA objective and directed to a spectrograph in the imaging mode. A typical recorded spectrum from another sample than the one depicted in (a) is shown in (c). The right bright region corresponds to the emission spectrum from the air gap at the end of the filled part, and the signal on the left is the spectrum of the scattered laser light from a single 120-nm gold nanoparticle.}
\end{figure}

We inserted gold nanoparticles (BBI Solutions) with a diameter of 120 nm into the capillaries prior to filling them with phenol. For this purpose, the water-suspended nanoparticles were transferred to methanol by centrifuging and solvent exchange. The glass capillary was then immersed into the suspension inside a glass test tube, and capillary forces helped filling it with the negatively charged nanoparticles. Next, the capillary was transferred directly into the phenol solution. Because of stronger capillary forces, phenol was sucked into the capillary and pushed the methanol further inside. Some of the nanoparticles were attached to the inner wall. Once the core was cooled down and solidified, we cleaned the exterior of the capillary. Finally, the filled region was cleaved, placed on a cover glass, and fixed with glue at the two end points.

The individual gold nanoparticles were imaged by dark-field microscopy as shown by the example in Fig.~\ref{fig:fiberside}(a). We used a grating and a camera to record the emission spectrum at each position along the fiber (see Fig.~\ref{fig:fiberside}(b)). Figure~\ref{fig:fiberside}(c) displays an example of the spatio-spectral images recorded from the laser action along such a capillary. For this experiment, the sample was pumped from the side by using a cigar-shaped beam. The pump beam cross section was chosen to be several times larger than the measurement field of view in order to provide a homogeneously excited region although this resulted a lower pumping efficiency.

At low pump intensities, a weak spontaneous emission could be seen along the core of the capillary, while a spectral narrowing at higher pump powers was observed in the light scattered from the air gap at the end of the filled region. This clearly demonstrates that the laser emission is fully coupled to the guided mode at the core and the scattering of the guided light from the imperfections at the core is negligible. For each shot, the same spectral features were detected in scattering from the gold nanoparticles inside the mode, albeit with less intensity. The center of emission spectrum is slightly shifted relative to the first experiment. This shift can possibly be due to the mixing of the solvents since liquid phenol has a high solubility in methanol. We also observed sharp peaks, which in parts might be due to the cavity modes that are formed between the gold nanoparticles and the end of the filled region. However, with the current experimental settings, we cannot test this hypothesis with an independent measurement.

\subsection{\label{sec:fine}Spectral fine-structure}

\begin{figure}[t]
\includegraphics[width=16cm]{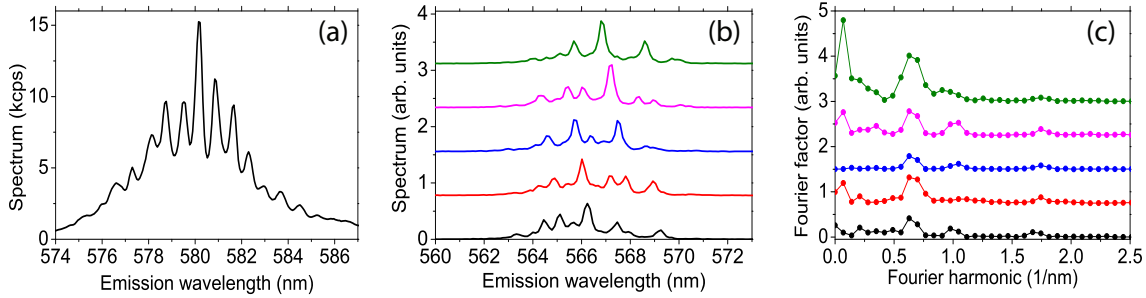}
\caption{\label{fig:finespectra} (a) Central part of the laser emission spectrum from the same sample as in Fig.~\ref{fig:threshold}. The pumping power was 4 times above the laser threshold and the spectra were averaged over 200 single shots. The regular separation between fringes amounts to $0.77$~nm. The visibility of these fringes is enhanced at higher pump intensities. (b) Five single-shot spectra taken from the side of a capillary containing extra gold nanoparticles, as illustrated in Fig.~\ref{fig:fiberside}. These peaks shift at each shot, possibly after melting of the material around the gold nanoparticle due to heating, but their relative spacings remain roughly the same. (c) Discrete Fourier transform of the spectra shown in (b).}
\end{figure}

We now turn our attention to the details of the emission spectra presented earlier. To analyze these further, we performed discrete Fourier transform (DFT) on several single-shot recorded spectra to distinguish possible resonant features and their effective lengths, similar to the reported analysis of random laser emission~\cite{polson_spectral_2000}. For the capillaries that were attached on top of the glass slides from one end, we found equally spaced fringes on top of a narrow laser emission, as depicted in Fig.~\ref{fig:finespectra}(a). We attribute the observed fringe spacing of $0.77$~nm to the Fabry-Per\'{o}t interference caused by the standard 170~$\mu$m thickness of the substrate glass slide with a refractive index of 1.51. This intriguing observation indicates the ultrahigh sensitivity of this microlaser to interfaces.

To confirm this interpretation, we performed a control experiment, in which we kept the capillaries inside the immersion liquid (water) and far apart from the glass substrate. The aforementioned fringes were no longer stable, and the intensities of the narrow spikes varied substantially from shot to shot. This behavior is a well-known aspect of nRFLs and has been also discussed in previous works on random lasers~\cite{el-dardiry_experimental_2012}.

We also analyzed the fine-structure of the emission from lasers with gold nanoparticles. Examples of these single-shot spectra are presented in Fig.~\ref{fig:finespectra}(b). The spectral position and the intensity of observed spikes varies substantially between shots. Despite these fluctuations, the spacing between the spectral fringes is rather stable, as can be seen from the DFT of each shot presented in Fig.~\ref{fig:finespectra}(c). The variations are most probably due to displacement of the nanoparticle. The persistent Fourier peak at 0.65/nm corresponds to a passive cavity length of 68~$\mu$m considering the refractive index of phenol, which is different from the spacing between the nanoparticle and the end of the filled region. A better understanding of this observation requires further investigation. We point out that in these first experiments, making samples with nanoparticles inside the gain medium has been quite difficult and has provided a low yield. The high concentration of the dye makes the formation of a stable colloidal suspension a challenge even in polar solvents. Furthermore, accessing the inner surface of capillaries is limited to the two ends and hence chemical modification of the inner surface becomes a nontrivial task.

\section{Discussion}
Lasing in nanofibers at high $\beta$-factors paves the way for low-threshold fluidic microlasers that can be integrated into opto-fluidic devices. Additionally, lasers with low temporal coherence, as demonstrated here, have been suggested for speckle-free imaging~\cite{redding_speckle-free_2012}. The high $\bfactor$ of our system has also made it possible to couple out the laser emission by a plasmonic nanoparticle. In future, we plan to tune the microlaser mode by controlling the cavity length through an active manipulation of the particle position, e.g. by applying electrostatic forces or using optical tweezers. Furthermore, our devices can be used as model systems for studying quantum behavior in other forms of collective emission such as superradiance~\cite{dicke_coherence_1954} and superfluorescence~\cite{bonifacio_cooperative_1975}.

The platform of a nanofiber laser can also serve as a simplified geometry for studying nRFLs and random lasers~\cite{ge_quantitative_2008}. Nonresonant feedback laser action is a fundamentally intriguing phenomenon and its theoretical modeling has been the subject of various and sometimes opposing interpretations~\cite{zaitsev_recent_2010, el-dardiry_experimental_2012}. A common feature of these lasers is the presence of strongly damped passive modes, which require a dynamic redefinition of the mode structure as a function of pump intensity due to gain competition and spatial hole-burning~\cite{tureci_strong_2008}. Sustaining the necessary population inversion for laser action in such lossy cavities would only be possible with a highly concentrated gain medium and large pump rates using pulsed excitation. Under such conditions, a relatively small feedback would suffice to select the laser modes based on their thresholds~\cite{goetschy_euclidean_2011}. Lasing is, thus, very sensitive to small variations in the medium so that the emission spectrum and the intensities of its components may fluctuate.

\section*{Acknowledgements}
We thank Michael Frosz at the Fiber Fabrication Technology Development and Service Unit of MPL for providing the capillaries, Andreas Br\"auer of the School of Advanced Optical Technologies (SAOT) for the generous loan of a pulsed laser and Harald Haakh for stimulating discussions. S. F. acknowledges fruitful discussions with Martin van Exter and Sergey Skipetrov and thanks Michel Orrit for his hospitality at Leiden University during the writing process of this manuscript.

\section*{References}
\bibliographystyle{ieeetr}
\bibliography{NJPreferences}

\end{document}